\documentstyle[12pt,epsfig]{article}
\def\be{\begin{equation}}
\def\ee{\end{equation}}
\def\ba{\begin{eqnarray}}
\def\ea{\end{eqnarray}}

\def\fun#1#2{\lower3.6pt\vbox{\baselineskip0pt\lineskip.9pt

\ialign{$\mathsurround=0pt#1\hfill##\hfil$\crcr#2\crcr\sim\crcr}}}
\begin{document}

\begin{titlepage}
\begin{flushright} 
hep-ph/9705210 \\
April 1997
\end{flushright}
\vspace{0.1in}

\centerline{{\Large \bf Large Top Quark
Yukawa Coupling}}
\centerline{{\Large \bf and Horizontal Symmetries}}

\vspace{0.35in}
\centerline{\large \bf Andrija Ra\v{s}in }
\vspace{0.15in}
\centerline{\it High Energy Section, International Centre for 
Theoretical Physics} 
\centerline{\it Strada Costiera 11, 34100 Trieste, Italy}
\vspace{0.5in}
\baselineskip=19pt

\centerline{\Large \bf Abstract}
\begin{quotation}

We consider the maximal U(3) horizontal scheme as a handle on fermion
masses and mixings. In particular, we attempt to explain the large top
Yukawa coupling and the masses and mixing in the two heaviest
generations. A simple model is constructed by enlarging the matter 
content of the Standard Model with that of a $10+\overline{10}$ 
pair of SU(5). 

The third generation particles get their masses when U(3) is broken 
to U(2). Top quark mass is naturally of order one. Bottom and tau 
masses are suppressed because of a hierarchy in the effective Yukawa
couplings and {\it not} from the hierarchy in the Higgs doublet
vacuum expectation values. The hierarchy is a consequence of the
fact that the particle spectrum contains an incomplete vector-like
generation and can come from hierarchies between scales of breaking of
different grand unified groups. 

Hierarchies and mixings between the second and third generation
are obtained by introducing a single parameter $\epsilon'$
representing the breaking $U(2) \rightarrow U(1)$.
As a consequence, we show that the successful (and previosuly
obtained) relations $V_{cb} \approx {m_s \over m_b} 
\approx \sqrt {m_c \over m_t}$ easily follow from our scheme.



\end{quotation}
\end{titlepage}

\baselineskip=19pt

\section{Introduction}

Despite all its successes, the Standard Model (SM) has
many unexplained features. Most of them are connected
to the fermionic sector, like the puzzling pattern of
masses and mixings or the fact that quarks and leptons
seem to neatly fit into three identical generations.
The situation is best summarized by the fact that of
the 19 arbitrary parameters in SM, 13 reside in the
fermionic sector. Thus, it seems that the search for a 
way beyond SM will go through the reduction of arbitrary
parameters in the fermionic sector.

A promising approach to explain some of SM features 
is that of using the flavor symmetry
of gauge interactions of the fermions, which is the
$U(3)^5$ global symmetry of rotations with each $U(3)$ belonging to one of
the five charged fermion sectors of SM ($q$,$u^c$,$d^c$,$l$,$e^c$).
This flavor symmetry is broken by various degrees by
the arbitrary Yukawa couplings of SM.
The idea essentially amounts to building an extension
of SM that is invariant under a certain
subgroup of the maximal flavor symmetry with Yukawa couplings
generated only when this horizontal symmetry gets broken.
SM is then the effective theory with Yukawa couplings carrying the 
information on the broken horizontal symmetry.
An example of this approach is the Froggatt-Nielsen mechanism (FN)
\cite{frog79,bere83}, in which the Yukawa couplings get generated from
higher dimensional operators when new scalars $\phi$, {\it flavons},
get their vacuum expectation values (VEVs)
and break the horizontal symmetries. 
The higher dimensional operators itself get generated by integrating out
some extra matter or scalar fields with mass M (for example
see later Figures 4 and 5).

Since the top quark mass is of order weak scale\cite{fabe95}, its 
corresponding Yukawa coupling is of order unity.  On the other
hand if the VEVs of the Higgs doublets in the theory are comparable, 
b and $\tau$ couplings are much smaller than one.
In this paper we build a model that incorporates a
large top and small bottom and tau Yukawa couplings.

How does one include the large top quark Yukawa coupling in a horizontal
symmetry model? In theories based on Abelian symmetries or SU(2),
it is usually assumed that the top does not transform under the
horizontal symmetry considered, noting that it must come from a maximally
broken SU(3).

In the maximal horizontal group SU(3) it does not
make sense to say that such a large number comes from a higher 
dimensional operator which is suppressed by some inverse powers of some 
high scale M. Rather it means the horizontal symmetry in 
the top quark sector is broken maximally, i.e. the VEV of the $\phi$ is of
the order of M. This means that if the horizontal 
symmetries were operative once at some high scale, either the third 
generation would have some large, unsuppressed 
mixing to some extra matter (unlike the other lighter generations)
or the SM Higgs doublet is maximally mixed with some new
scalars. In this paper we present a model which explores the first
possibility.

We consider the full global U(3) symmetry in a manner similar
to the U(2) case of Ref. \cite{barb95,barb96}. 
U(2) (or SU(2)) horizontal symmetry
has received lot of interest lately as a natural solution to the 
SUSY flavor problem, forcing the squarks of the first two families 
to be approximately degenerate\cite{dine93}. 
Thus, we will consider supersymmetric 
theories although we focus on conclusions in the fermionic sector
(we discuss the scalar sector briefly at the end).
The different hierarchies between $m_c / m_t$  and $m_s / m_b$ are 
also easily explained, as well as $V_{cb}$.
The large top quark 
mass is explained by the addition of an extra 
$10+\overline{10}$ of SU(5) \footnote {A similar field content in the
context of supersymmetry was also recently proposed by Berezhiani in
\cite{bere96}.}. 
The theory can also explain the smallness of the bottom and tau lepton
masses without any suppression of Higgs doublet VEVs.  
It is easy to accommodate also the first generation in this scheme, 
but we chose to avoid doing so in this paper
for clarity of the argument and reasons we discuss later. 

Motivated by grand unification and more predictivity, we consider 
the same U(3) acting on
all charged sectors (rather than the maximal $U(3)^5$). 
Thus the scale M could be  
some scale of order $10^{16}$ GeV or so, although we will comment
on how low phenomenologically such a scale can be. 

The feature of large mixings of the top with extra matter was explored 
in several papers. 
For example, in a supersymmetric Pati-Salam 
model there is an extra gaugino with charge $+2/3$ which can effectively play
the role of an extra vector-like quark singlet, and the large top quark
mass can be related to the scale of SUSY breaking which is of the order of
the weak scale\cite{barb89,babu90a}. 
Many other papers explore the possibility of having an extra 
vector-like singlet up quark\cite{bala88}.
The issue of
large top Yukawa coupling and large mixing in an inverse hierarchy
scheme was discussed in Ref. \cite{bere93}. A pseudogoldstone approach
for the Higgs doublets where the top mixes with extra vector-like matter
can be found in Ref. \cite{barb94}.

First attempt at building supersymmetric theories with nonabelian
horizontal symmetries was done by Berezhiani {\it et al.} in
\cite{bere86}. Later attempts include those listed in \cite{poul93}.
Cosmological consequences of a global SU(3) family symmetry broken at a GUT scale
were studied in \cite{joyc94}.

We start in Section 2 with the masses of the third generation. We show how the
top quark Yukawa coupling can be generated from the breaking of U(3)
and still be of order one, while the bottom and tau Yukawa
couplings are suppressed {\it without} a hierarchy in the
VEVs of the standard Higgs doublets. The masses of the second generation
fermions, discussed in Section 3,  are generated in a 
manner somewhat similar to
reference \cite{barb96} and 
come from the breaking of the remaining U(2) 
symmetry down to U(1). Section 4 is reserved for the discussion of the
origin of the nonrenormalizable terms and the generated ratio $m_b \over m_t$.
We conclude with some final thoughts in Section 5.


\section{Mass of the third generation fermions}

In order to explain the large top quark Yukawa coupling within a
FN scheme we must add some extra matter fields. There
exist strong limits on extra matter, like SM-like generations, from
electroweak precision measurements\cite{pdg96}. However, extra vector-like
matter is almost not constrained. Furthermore, gauge coupling 
unification is not spoiled if matter is added in 
$5+\overline{5}$ or $10+\overline{10}$s\footnote{Perturbativity however
constrains
the number of such extra pairs\cite{moro93}.}.
It is interesting that string compactification can give three generations
and extra
vector-like matter with a SM invariant mass which is not 
necessarily at the Planck scale\cite{gree86,maal90}.  
Since we will discuss the grand unification of such a theory,
we will assume masses of the order GUT scale, although we will
comment later on how low can such a scale be.

We add to the three generations of the SM 
($q_a$, $u^c_a$, $d^c_a$, $l_a$, $e^c_a$, $a=1,2,3$)
vector-like matter with the content of $10 + \overline{10}$ 
of SU(5)
\be
\begin{array}{ccc}
Q_1 & U^c_1 & E^c_1  \\
Q^c_2 & U_2 & E_2 \, .
\end{array}
\ee
Index $a$ denotes generations and goes from 1 to 3. 
Notice that 
the $SU(2)_L$ \underline{doublet}
$Q^c_2$ carries fields with the same electric charges as the $SU(2)_L$
\underline{singlets} $u^c_a$ and $d^c_a$, and that the $SU(2)_L$
\underline{singlet} $U_2$ carries the same electric charge
as the up quark fields in the $SU(2)_L$ \underline{doublets} $q_a$.
Also, the $SU(2)_L$ \underline{singlet} $E_2$ carries the same electric charge
as the charged field in the $SU(2)_L$ \underline{doublets} $l_a$.
We will call ``mismatch" this wrong pairing of 
fields from different SU(2) multiplets with the same electric charge.
As is well known and we show later, this will make the quark mixing matrix 
non-unitary.

{\bf Mass of the top}

We assume that the horizontal group is maximal, i.e. U(3), under which
the three generations transform as $\underline{3}$, while the extra 
vector-like
matter is neutral. In addition, we assume that there is an extra SM singlet
flavon field $\phi^a$ which transforms as a $\underline{\overline{3}}$ under 
U(3), 
and the upper index denotes a charge opposite to the charge of a field with
a lower index.
We denote the gauge invariant mass of the extra vector-like matter by M.

Thus the most general mass terms come from  
\ba
\phi^a Q^c_2 q_a + \phi^a u^c_a U_2 + \phi^a e^c_a E_2 +
M Q^c_2 Q_1 \nonumber\\
+ M U^c_1 U_2 + M E^c_1 E_2 + H_2 U^c_1 Q_1 + H_1 Q^c_2 U_2 \, ,
\label{firstterm}
\ea
from which we obtain the mass matrices for 
up quarks, down  quarks and charged 
leptons

\be
L_{M_u} = (  u^c_a \, U^c_1 \, U^c_2 ) \, 
\left(
\begin{array}{ccc}
{\bf 0} & {\bf 0}  & \phi^a \\
{\bf 0} & H_2^0 & M \\
\phi^a & M & H_1^0
\end{array}
\right)
\,  
\left( 
\begin{array}{c}
u_a \\
U_1 \\
U_2
\end{array}
\right)
+ {\rm h.c.} \, ,
\label{yukt}
\ee

\be
L_{M_d} = ( d^c_a \, D^c_2 ) \, 
\left(
\begin{array}{cc}
{\bf 0} & {\bf 0}\\
\phi^a & M
\end{array}
\right)
\,  
\left( 
\begin{array}{c}
d_a \\
D_1 
\end{array}
\right)
+ {\rm h.c.} \, ,
\label{yukb}
\ee

\be
L_{M_e} = ( e^c_a \, E^c_1 ) \, 
\left(
\begin{array}{cc}
{\bf 0} & \phi^a \\
{\bf 0} & M
\end{array}
\right)
\,  
\left( 
\begin{array}{c}
e_a \\
E_2 
\end{array}
\right)
+ {\rm h.c.} \, ,
\label{yuke}
\ee
where $H_1$ and $H_2$ are the Higgs doublets.
Index $a$ runs from 1 to 3 so that the up type mass matrix is a $5 \times 5$,
while the down and lepton matrices are $4 \times 4$. Boldfaced zeroes denote
the appropriate matrix, vector or column with all elements equal to zero.
Also, notice that, for example, $U^c_2$ from the doublet $Q^c_2$ is 
grouped with the singlets, reflecting the mismatch.
For simplicity, we assumed that the up
quark mass matrix is symmetric\footnote{For example, the gauge invariant
mass terms $U^c_1 U_2$ and $U^c_2 U_1$ do not have to be exactly equal. 
However this does not qualitatively change our results.}.

The VEV of $\phi$ can always be rotated so that only one component
obtains a VEV, say $\phi^3$. Thus, $<\phi^3>$ breaks the U(3) symmetry 
down to U(2). If we diagonalize the up quark mass matrix (\ref{yukt})
we get the top quark with mass 
\be
m_t = v_2 { {<\phi^3>^2} \over {<\phi^3>^2 + M^2} } \, ,
\label{topmass}
\ee
while $b$ and $\tau$ remain massless
\be
m_b = m_\tau = 0.
\ee
Equation (\ref{topmass})
holds {\it regardless} of the values
of $<\phi^3>$ and $M$, 
as long as they are both larger than $v_1$ and 
$v_2$. In particular, we take 
$<\phi^3>$ and $M$ of the same order, in order for the top 
quark mass to be of the order weak scale \cite{babu90}.
In addition, there are four heavy states with mass 
$\sqrt{<\phi^3>^2+M^2}$ (two in the up sector, one in down and
one in lepton sector).

Let us discuss the possibility that $<\phi>$ (scale of maximal horizontal
symmetry breaking) is of order M (mass of the vector-like pair). In fact,
there is {\it a priori} nothing that can stop it from doing so: the scalar potential
involving $\phi$ will have all dimensionful parameters at scale M.
Only subgroups
that are preserved after the breaking of the largest group may get
a smaller VEV than the natural scale M at some later stage (possibly from 
radiative corrections).

The grand unified origin of terms in (\ref{firstterm})
is straightforward. 
We consider a Froggatt-Nielsen (FN) \cite{frog79,bere83} 
theory with the flavor group U(3) and unification group SU(5).
Ordinary matter is in $t_a (10) + f_a (\overline{5})$, extra
FN vector matter is in $T (10) + \overline{T} (\overline{10})$, 
Higgs fields are in H(5) and $\overline{\rm H}(\overline{5})$, where the 
transformation properties under SU(5) are spelled out in the brackets. 
Flavons $\phi^a$ are SU(5) singlets. The most general renormalizable
interactions are
\be
\phi^a \overline{T} t_a + 
M \, T \overline{T} + 
T \, T \, H + \overline{T}
\, 
\overline{T} \overline{H} \, .
\ee
On integrating out heavy states, there is a single diagram, given
in Fig. 1, which generates the top mass as in (\ref{topmass}).
\begin{figure}[ht]
\epsfig{file=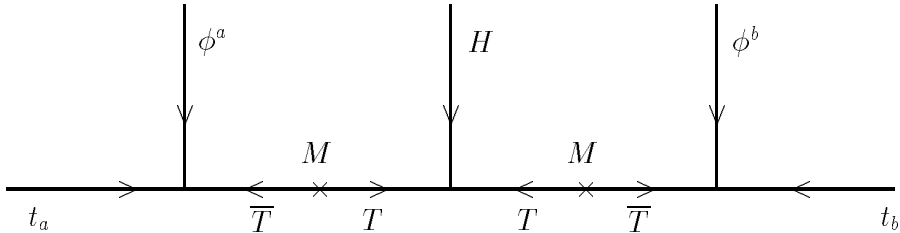,width=12cm}
\caption{Mechanism for generating the top quark mass.}
\end{figure}
We conclude that the top is heavy because at the scale of U(3) breaking
the only FN fields transform as 10.

{\bf Mass of the bottom and tau}

Masses of lighter fermions may be generated in a way similar to the
U(2) case \cite{barb96}. We use a flavon field $\phi^{ab}$ 
which is symmetric in flavor indices 
(a $\underline{6}$ of U(3)) and which
can generate some higher order operators of the form 
\be
{ {\phi^{ab}} \over M_H} [ u^c_a q_b H_2 + d^c_a q_b H_1 + e^c_a l_b H_1 ]
\, .
\label{bottau}
\ee
The crucial point is that the mass $M_H$ has no reason to be of the same 
order as M. Mass M is the SU(5) invariant mass of the 
$10+\overline{10}$ pair,
while $M_H$ can for example come from the SO(10) invariant mass, and thus 
can be higher by several orders of magnitude. This will be discussed in
more details in Section 4.
From now on, we will assume 
\be
\epsilon \equiv M / M_H \approx 10^{-2} \, .
\label{assume}
\ee
The field $\phi^{ab}$ may be a new field added to the theory or, more
economically, an effective field made out of the product of two 
fundamentals $\phi^a\phi^b$.
Later we will give an explicit SU(5) realization of the model
where we discuss possible ways of generating the hierarchy (\ref{assume}).

U(3) is also broken by the VEV of $\phi^{33}$ and we expect it to be
of the 
same order as $<\phi^3> \approx M$. This then modifies the Yukawa 
matrices in 
(\ref{yukt})-(\ref{yuke}) in the 3,3 entry
\be
L_{M_u} = (  u^c_i \, u^c_3 \, U^c_1 \, U^c_2 ) \, 
\left(
\begin{array}{cccc}
{\bf 0} & {\bf 0} & {\bf 0} & {\bf 0} \\
{\bf 0} & { <\phi^{33}> \over M_H } v_2  & 0 & <\phi^3> \\
{\bf 0} & 0 & v_2 & M \\
{\bf 0} & <\phi^3> & M & v_1
\end{array}
\right)
\,  
\left( 
\begin{array}{c}
u_i \\
u_3 \\
U_1 \\
U_2
\end{array}
\right)
+ {\rm h.c.} \, ,
\ee

\be
L_{M_d} = ( d^c_i \, d^c_3 \, D^c_2 ) \, 
\left(
\begin{array}{ccc}
{\bf 0} & {\bf 0} & {\bf 0} \\
{\bf 0} & { <\phi^{33}> \over M_H } v_1  & 0 \\
{\bf 0} & <\phi^3> & M
\end{array}
\right)
\,  
\left( 
\begin{array}{c}
d_i \\
d_3 \\
D_1 
\end{array}
\right)
+ {\rm h.c.} \, ,
\ee

\be
L_{M_e} = ( e^c_i \, e^c_3 \, E^c_1 ) \, 
\left(
\begin{array}{ccc}
{\bf 0} & {\bf 0} & {\bf 0} \\
{\bf 0} & { <\phi^{33}> \over M_H } v_1 &  <\phi^3> \\
{\bf 0} & 0 & M
\end{array}
\right)
\,  
\left( 
\begin{array}{c}
e_i \\
e_3 \\
E_2 
\end{array}
\right)
+ {\rm h.c.} \, ,
\ee
where $i=1,2$.
Diagonalizing we see that
the top quark mass stays almost unchanged. However,
bottom and tau masses are generated and they are of order 
\be
m_b \approx m_{\tau} 
\approx  { <\phi^{33}> \over M_H } v_1 
\approx {M \over M_H} v_1 
\equiv \epsilon \, v_1 
\, .
\label{btau}
\ee

This realization of the heaviest generation masses is different than
\cite{bere96}, where the top-bottom splitting was left to be
explained as usual (either a
large ratio of the Higgs doublet VEVs (large $\tan\beta$) or a
large ratio of Yukawa couplings put in by hand).

The terms in (\ref{bottau}) can be generated in SU(5) 
if we add the nonrenormalizable operators suppressed by the
high scale $M_H$
\be
{ \phi^{ab}  \over M_H} ( t_a t_b H + 
t_a \overline{f}_b \overline{H}) \, ,
\label{symmop}
\ee
as shown in the diagram of Fig. 2. 
\begin{figure}[ht]
\epsfig{file=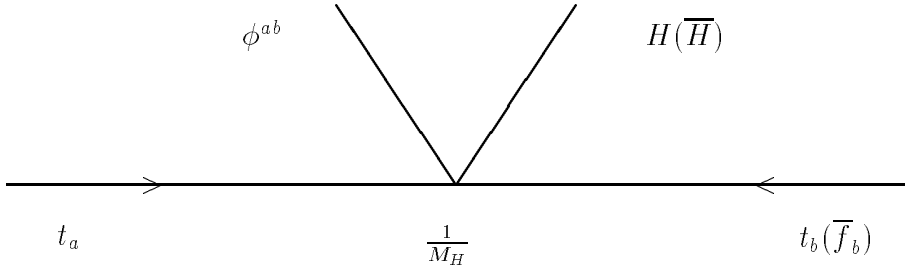,width=12cm}
\caption{A mechanism for generating bottom and tau masses. 
This mechanism can also be used to generate lighter generation
masses.}
\end{figure}
Instead of using the symmetric flavon field $\phi^{ab}$ 
to generate the bottom and tau masses, a  more
economical way is to use only the fundamental fields $\phi^a$. 
In this case we introduce nonrenormalizable operators of the form
\be
{1 \over M_H} ( T \overline{f}_a \phi^a \overline{H}
+ T t_a \phi^a H ) \, ,
\label{econop}
\ee
which generate nonzero entries in the up, down and lepton mass matrices
from the diagrams shown in Fig. 3.
\begin{figure}[ht]
\epsfig{file=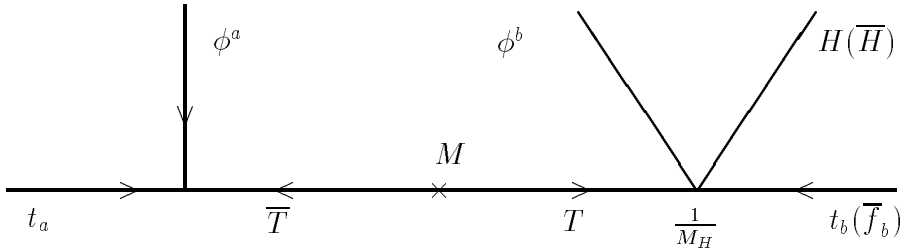,width=12cm}
\caption{Another mechanism for generating bottom and tau masses.}
\end{figure}

Before going on to generate masses of lighter generations, let us discuss
the diagonalization of the sector involving the third family and the extra 
vector like fields and the ensuing quark mixing matrix. 
It is obvious from the above equations that the
rotation to get the top and bottom mass involves the same rotation on the
left fields so that the KM matrix  element $V_{33}$ will be close to one.
Let us do this in more detail. 
The up quark mass matrix is 
diagonalized by the following rotations both on the left and on the right:
\ba
V^u_L & = &  
V^u_R \approx  
\left(
\begin{array}{cccc}
{\bf 1} & {\bf 0} & {\bf 0} & {\bf 0} \\
{\bf 0} & {M \over {\sqrt{M^2 + <\phi^3>^2}} }  
& {<\phi^3> \over {\sqrt{M^2 + <\phi^3>^2}} }  & 0 \\
{\bf 0} & -{<\phi^3> \over {\sqrt{M^2 + <\phi^3>^2}} }  
& {M \over {\sqrt{M^2 + <\phi^3>^2}} }  & 0 \\
{\bf 0} & 0 & 0 & 1
\end{array}
\right)  \times \nonumber\\
& \times &
\left(
\begin{array}{cccc}
{\bf 1} & {\bf 0} & {\bf 0} & {\bf 0} \\
{\bf 0} & 1 & 0 & O({v_2 \over M}) \\
{\bf 0} & 0 & 1 & 0 \\
{\bf 0} & O({v_2 \over M}) & 0  & 1
\end{array}
\right)
\left(
\begin{array}{cccc}
{\bf 1} & {\bf 0} & {\bf 0} & {\bf 0} \\
{\bf 0} & 1 & 0 & 0 \\
{\bf 0} & 0 & c & s \\
{\bf 0} & 0 & -s  & c
\end{array}
\right) \, ,
\label{leftupmix}
\ea
while the down and charged lepton mass 
matrices are diagonalized by \be
V^d_L =  
V^e_R \approx  
\left(
\begin{array}{ccc}
{\bf 1} & {\bf 0} & {\bf 0}  \\
{\bf 0} & {M \over {\sqrt{M^2 + <\phi^3>^2}} }  
& {<\phi^3> \over {\sqrt{M^2 + <\phi^3>^2}} }  \\
{\bf 0} & -{<\phi^3> \over {\sqrt{M^2 + <\phi^3>^2}} }  
& {M \over {\sqrt{M^2 + <\phi^3>^2}} }
\end{array}
\right) \, ,
\label{leftdownmix}
\ee
up to corrections of order ${v^2 \over M^2}$ or ${{\epsilon \, v} \over M}$. 
The last matrix on the right hand side of equation (\ref{leftupmix})
is the rotation in the heavy sector and is unimportant for our
discussion\footnote{$s = {1 \over \sqrt{2}}$ for symmetric up matrix, and 
$s=O({v \over M})$ otherwise.}. From the above discussion we see that
the lefthanded and righthanded top, lefthanded bottom and righthanded tau
are actually maximally mixed states (for $<\phi> \approx M$) 
of the third flavored generation and the extra matter
in 10 of SU(5).

In the Appendix we show that the quark mixing matrix is in fact a 
$10\times8$
matrix with two $5 \times 4$ blocks. The upper block contains the 
light quark mixings in the $3 \times 3$ sector which can be identified
with the Kobayashi-Maskawa (KM) matrix 
$\sum_{k=1,4} (V^{u\dagger}_L)_{ik} 
(V^d_L)_{kj}$ (i,j=1,2,3).
We see at this level, from (\ref{leftupmix}) and (\ref{leftdownmix}),
that the KM matrix is equal to unity. 
Departure of elements from those of the unit matrix is of the order 
${v^2 \over M^2}$ or ${{\epsilon v} \over M}$
which is negligible for M of order GUT or Planck 
scale.

In the lepton sector the situation is similar. The right handed
rotation defines a combination of $e^c_3$ and $E^c_1$ as the
righthanded component of the tau lepton. 

A note on scalar masses. It is interesting that the 
righthanded down squarks and lefthanded sleptons
remain approximately degenerate even though
the U(3) symmetry is broken\cite{bere96}. This is because the rotations 
on these fields are suppressed by $\epsilon$. We mention the consequences
of this towards the end of the paper.

\section{Mass of the second generation fermions}

Second generation masses are generated when the remaining U(2) symmetry
breaks down to U(1) (which keeps the first generation massless). 
We can obtain this breaking economically from the same symmetric flavon field 
$\phi^{ab}$ (or an additional symmetric field $\phi'^{ab}$), when it gets 
VEVs of the same order in the (2,2), (2,3) and (3,2) entries.
We parametrize this breaking by a parameter 
\be
\epsilon' 
\equiv { <\phi^{22}> \over M_H }
\approx { <\phi^{23}> \over M_H }
= { <\phi^{32}> \over M_H } \, . 
\ee
The structure of the fermion mass matrices 
in the weak eigenstate basis is

\be
L_{M_u} = (  u^c_1 \, u^c_2 \, u^c_3 \, U^c_1 \, U^c_2 ) \, 
\left(
\begin{array}{ccccc}
0 & 0 & 0 & 0 & 0 \\
0 & \epsilon' v_2 & \epsilon' v_2 & 0 & 0 \\
0 & \epsilon' v_2 & \epsilon  v_2  & 0 & <\phi^3> \\
0 & 0 & 0 & v_2 & M \\
0 & 0 & <\phi^3> & M & v_1
\end{array}
\right)
\,  
\left( 
\begin{array}{c}
u_1 \\
u_2 \\
u_3 \\
U_1 \\
U_2
\end{array}
\right)
+ {\rm h.c.} \, ,
\label{totmassu}
\ee

\be
L_{M_d} = ( d^c_1 \, d^c_2 \, d^c_3 \, D^c_2 ) \, 
\left(
\begin{array}{cccc}
0 & 0 & 0 & 0 \\
0 & \epsilon' v_1 & \epsilon' v_1  & 0 \\
0 & \epsilon' v_1 & \epsilon v_1  & 0 \\
0 & 0 & <\phi^3> & M
\end{array}
\right)
\,  
\left( 
\begin{array}{c}
d_1 \\
d_2 \\
d_3 \\
D_1 
\end{array}
\right)
+ {\rm h.c.} \, ,
\label{totmassd}
\ee

\be
L_{M_e} = ( e^c_1 \, e^c_2 \, e^c_3 \, E^c_1 ) \, 
\left(
\begin{array}{cccc}
0 & 0 & 0 & 0\\
0 & \epsilon' v_1 & \epsilon' v_1 &  0 \\
0 & \epsilon' v_1 & \epsilon v_1 &  <\phi^3> \\
0 & 0 & 0 & M
\end{array}
\right)
\,  
\left( 
\begin{array}{c}
e_1 \\
e_2 \\
e_3 \\
E_2 
\end{array}
\right)
+ {\rm h.c.} \, .
\label{totmasse}
\ee

Diagonalizing the third generation+heavy sector as in the previous section
(with $<\phi^3> \approx M$)
will not change much the structure of the second generation sector
\footnote{Compact formulas for block diagonalizing such matrices
can be found in the Appendix of \cite{babu95}.}:

\be
L_{M_u} = (  u^c_1 \, u^c_2 \, u^c_{M3} \, u^c_{M4} \, u^c_{M5} ) \, 
\left(
\begin{array}{ccccc}
0 & 0 & 0 & 0 & 0 \\
0 & \epsilon' v_2 & \epsilon' v_2  & (\epsilon' v_2) & (\epsilon' v_2) \\ 
0 & \epsilon' v_2  & v_2  & 0 & 0 \\
0 & (\epsilon' v_2) & 0 & M & 0 \\
0 & (\epsilon' v_2) & 0 & 0 & M 
\end{array}
\right)
\,  
\left( 
\begin{array}{c}
u_1 \\
u_2 \\
u_{3M} \\
u_{4M} \\
u_{5M}
\end{array}
\right)
+ {\rm h.c.} \, ,
\label{diagup2}
\ee

\be
L_{M_d} = ( d^c_1 \, d^c_2 \, d^c_{M3} \, d^c_{M4} ) \, 
\left(
\begin{array}{cccc}
0 & 0 & 0 & 0 \\
0 & \epsilon' v_1 & \epsilon' v_1  & 0 \\
0 & \epsilon' v_1  & \epsilon v_1  & 0 \\
0 & (\epsilon' v_1) & 0 & M
\end{array}
\right)
\,  
\left( 
\begin{array}{c}
d_1 \\
d_2 \\
d_{3M}\\
d_{4M} 
\end{array}
\right)
+ {\rm h.c.} \, ,
\label{diagdown2}
\ee

\be
L_{M_e} = ( e^c_1 \, e^c_2 \, e^c_{3M} \, e^c_{4M} ) \, 
\left(
\begin{array}{cccc}
0 & 0 & 0 & 0\\
0 & \epsilon' v_1 & \epsilon' v_1 &  (\epsilon' v_1) \\
0 & \epsilon' v_1 & \epsilon v_1 &  0 \\
0 & 0 & 0 & M
\end{array}
\right)
\,  
\left( 
\begin{array}{c}
e_1 \\
e_2 \\
e_{3M} \\
e_{4M} 
\end{array}
\right)
+ {\rm h.c.} \, ,
\label{diaglepton2}
\ee
where we denoted only the order of magnitude of relevant entries,
and index $M$ denotes the approximate mass eigenstates.
We can neglect the bracketed terms (which represent mixings between the 
light and heavy fields) since they yield only  order 
${\epsilon' v \over M}$ mixings, without significantly changing mass 
eigenvalues. 

Note that the obtained structure of mass matrices
for the second and third generation is similar to the one of the U(2) 
model in Ref. \cite{barb96}.

Now, we see immediately that the following relations approximately hold:
\be
{m_c \over m_t} \approx \epsilon' \, ,
\ee
\be
{m_s \over m_b} \approx {m_\mu \over m_\tau} \approx 
{\epsilon' \over \epsilon} \, .
\ee
We see that for $\epsilon' \approx \epsilon^2$ we get good agreement
with experiment. Moreover, we get the successful 
relations \cite{barb96,mohaxx}
\be
V_{cb} \approx {m_s \over m_b}
\approx \sqrt {m_c \over m_t} \, .
\ee

These relations will also approximately hold in the SU(5) grand unified 
theory, with the relevant contributions coming from diagrams as the 
one shown in Figure 2. Although not the main aim of this paper, one can
make the relations more precise.
We assume that the VEVs of 
$\phi^{22}$ and $\phi^{23}$ 
are of order $\epsilon' M_H$ since they break U(2) down to U(1).
At first sight, the precise relations among
the four observables $m_c$, $m_s$, $m_\mu$ and $V_{cb}$ can
always be fixed to fit the experimental values by the 
four unknown numbers of order one:  
$<\phi^{22}> / (\epsilon' M)$, 
$<\phi^{23}> / (\epsilon' M)$ and 
the numbers of order one in front of the two nonrenormalizable
operators in (\ref{symmop}). However, if the flavon field $\phi^{ab}$
is a SU(5) singlet then we have the relation $m_\mu = m_s$ at
the GUT scale. However, notice that if $\phi^{ab}$ that contributes
to the (2,2) entry is such a 
multiplet that 
$\phi^{ab} \overline{H}$ is a $45$ of SU(5),
then a more successful relation emerges
at the GUT scale $m_\mu = 3 m_s$\cite{geor79}. 
For example, this can be achieved with
$\phi^{ab}$ in 24 or 75 of SU(5). 
However notice that this then forbids the up quark mass entries
(and thus $m_c$), since the
$45$ is in the antisymmetric part of $10 \times 10$,
thus prompting the use of more complicated representations. 
For example if one wants the $\phi^{ab}$ to lie in the 24, one can also
construct $\phi^{ab} \overline{H}$ as a $5$ of SU(5), with some additional
vector states (see next section). 

Although somewhat beyond the scope of this paper, it is possible to extend 
this analysis to explain the masses of the
first generation, using for example an antisymmetric representation of U(3). 
However, we chose not do so in this paper for clarity. Anyway 
the values of the lightest generation masses are still to some degree
undetermined because of at least two 
reasons. Planck scale physics can alter 
the values of the lightest fermions through higher dimensional operators.
Also the issue of whether the mass of the lightest quark is zero
or not is far from being settled\cite{leut94,choi88}.

\section{Origin of the nonrenormalizable operators and
the hierarchy $M << M_H$}

The nonrenormalizable operators with $\phi^{ab}$ introduced
in Section 3 can be generated in two ways from the 
heavy scale \cite{frog79,bere83,iban94}. One is the usual
Froggatt-Nielsen way in which there are heavy fields on
the matter line. For example, we can exchange a pair
of $T^a(10) + \overline{T}_a(\overline{10})$ with mass $M_H$ as shown in 
Figure 4. 
\begin{figure}[ht]
\epsfig{file=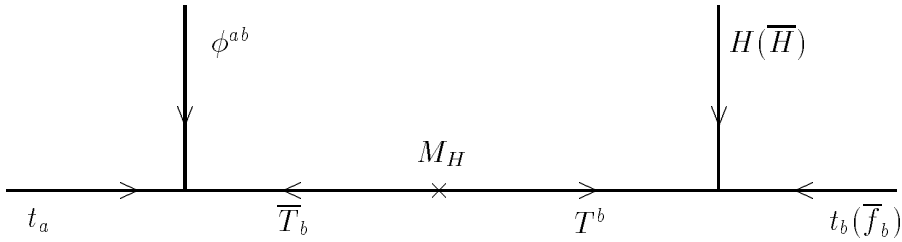,width=12cm}
\caption{A mechanism for generating the higher dimensional
operators in (\ref{symmop}).} 
\end{figure}
Another way\cite{iban94} is where there are some heavy fields on the 
line of the fields which get a VEV, as shown in Figure 5 (here the
$\phi^{ab}$ transforms as 24 under SU(5)).
\begin{figure}[ht]
\epsfig{file=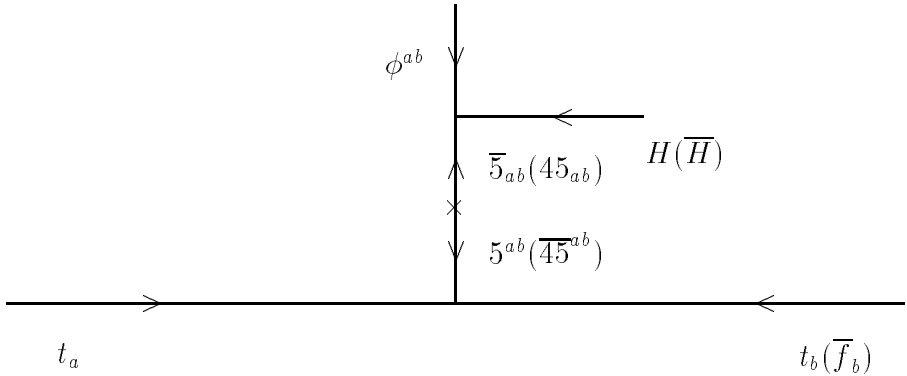,width=12cm}
\caption{Alternate mechanism for generating the higher dimensional
operators in (\ref{symmop}). } 
\end{figure}
The choice of 45 in Figure 5 is convenient to generate the 
desired relation $m_\mu = 3 m_s$ at the GUT scale.

Before going on one needs to explain the ratio of masses 
$M \over M_H$. One can explain easily such a ratio by a
discrete symmetry softly broken by the M mass term and
the VEVs of field $\phi^a$. However, there is also a deeper
understanding for such a ratio as we now explain. As advertised before, 
it is entirely possible that 
the origin lies in the different scales of breaking SU(5) and
SO(10).  Notice that $M$ is the SU(5) invariant mass 
of $T \overline{T}$, while $M_H$ is the SU(5) invariant 
mass of $T^a \overline{T}_a$. However, we could
have added also FN fields in the 
$F^a$(5) and  $\overline{F}_a (\overline{5})$ representations, also with
mass $M_H$, which would generate new higher dimensional operators,
but with no new contribution to the order of masses that
did not exist before. Thus, we can imagine that 
$T_a$ and $\overline{F}_a$ come from a $16_a$ of SO(10), and 
$\overline{T}_a$ and $F_a$ from a $\overline{16}_a$ of SO(10), 
so that $16_a$ and $\overline{16}_a$ can be combined to have an
SO(10) invariant mass $M_H$.
On the other hand new fields $F + \overline{F}$
are forbidden to have mass of order M because they would 
force the Yukawa couplings of the bottom and tau to
be of order one, in contrast to our assumption that the
Higgs doublet VEVs are of the order weak scale.
Thus mass $M$ is SU(5) invariant mass only of the pair
$T + \overline{T}$, while $F + \overline{F}$ mass remains
at the  higher scale $M_H$.

A similar way of understanding this is that the U(3) symmetry
is effectively a product of two symmetries 
$U(3) \times U(3)$. 
Imagine an SO(10) generalization of what we were doing so
far in SU(5), with $16$,$\overline{16}$,$16_a$, 
etc., with all mass scales
at $M_H$. The only thing we need is a 45 with a VEV (order $M_H$) in the 1
direction of SU(5). We need one fine tuning to make one pair of tens of
SU(5) light (mass M), while all other FN fields remain heavy (mass $M_H$).
This gives the effective SU(5) theory that we have. Now the U(3)
breaking will affect only the states in $t_a$, and is thus an effective
U(3) carried by the $t_a$. Breaking of the U(3) in the $\overline{f}_a$
sector is suppressed by $1/M_H$, and it looks like a separate U(3)
on those states. (For a recent work with product of SU(3)'s as horizontal
symmetries see \cite{anse97}).

\section{Final remarks and conclusions}

One can ask how low can the mass of the extra vector matter
fields be\cite{bowi83,senj84}. Suppose the mass matrices 
(\ref{totmassu})-(\ref{totmasse}) were given without resorting to 
horizontal symmetries. Then, interestingly enough, existing experimental
limits on $V_{tb}$ or flavor changing Z decays would allow M to be as low
as the weak scale, because all effects quickly decouple as M becomes
large\footnote{Present limits imply only $V_{tb} > 0.05$ or 
so\cite{gerd96}. Also oblique corrections quickly dissapear as the 
$SU(2)_L$ invariant mass M becomes larger\cite{lavo93,inam92}.}
(see Appendix).

However, a much stronger limit on M comes from the fact that we
are breaking global flavor symmetries spontaneously.
The Goldstone bosons, known as {\it familons}\cite{reis82,jkim87}, 
will actually produce too much FCNC, unless the scale is higher than
about $5 \times 10^{11}$ GeV \cite{atiy96}. 
The bound applies to the lowest of scales, $v$, in the chain
of scale of horizontal symmetry breaking 
$U(3) \stackrel{M}{\rightarrow} U(2) 
\stackrel{\epsilon \, M}{\rightarrow} U(1)
\stackrel{v}{\rightarrow} nothing$.
If we assume that $M$ lies near a typical GUT scale, and that
the lightest generation masses are generated by the last scale
in the chain, $v$ may live actually very close to the lower 
bound\cite{barb95}. Interestingly enough,
the symmetry being broken at the lowest scale, U(1), has a color anomaly,
so that we have an axion in the theory coming from the
family symmetry\cite{barb81}. Then there is also an upper
bound on $v$ as well coming from cosmology\cite{pres83}, 
$v < 10^{12}$ GeV or so. 

In this paper we envisioned the underlying theory to be a supersymmetric
one, although all conclusions presented here concern the fermionic
sector and are valid also in a nonsupersymmetric version. In SUSY,
the U(3) symmetry acts on scalar partners as well. Here we will
mention a few main points regarding the scalar masses and leave a 
more complete investigation to a future publication.
It is interesting that because of the choice of representations of 
extra matter (10+$\overline{10}$) right-handed 
squarks and left-handed sleptons remain approximately
degenerate in all three generations. This has some profound
differences compared to the recent analysis based on U(2). 
In comparison to \cite{barb95,barb96}, we expect
$\mu \to e \gamma$ and the electric dipole moment of the electron to be
suppressed by $\epsilon^2$ and the
$K-\bar{K}$ mass difference by $(m_s / m_K)^2$.
Recently, it
has been pointed out that if there is more than one operator responsible
for the same eigenvalue in the Yukawa matrix, 
the misalignment of A-terms and Yukawa terms
can actually produce the SUSY $\epsilon_K$ problem\cite{ynir96}. 
Our theory for lighter generations should essentially
come from the operator containing $\phi^{ab}$ (and possibly from an
antisymmetric flavon for the lightest generation) and is
thus of the same form as the U(2) models which relax this problem.
However, a more precise prediction of fermion masses and mixings
may require more fields (as alluded in Section 3). This then may require
proportionality in order to avoid the problem\cite{bere96}. 

\vspace{0.3in}

To summarize, in order for the Yukawa couplings of order one (top
and/or bottom) to find an explanation within the Froggat-Nielsen type 
of horizontal symmetry approach, it is necessary that the third generation 
particles mix maximally with some extra matter fields or that the 
Higgs doublet mixes maximally with extra scalars. In this paper we 
considered the first approach.

We have considered the maximal U(3) horizontal symmetry scheme,
with the emphasis on the two heaviest generations and
the large top Yukawa coupling. A simple scheme  can be achieved with the 
extra matter in $10+\overline{10}$ 
pair of SU(5). The third generation particles get their masses when
U(3) is broken to U(2). Top quark mass is naturally 
of order one. Bottom and tau masses 
are suppressed because of a hierarchy in effective Yukawa
couplings and {\it not} from the hierarchy in the Higgs doublet
VEVs. The hierarchy of effective Yukawa couplings can come from
hierarchies between scales of breaking of different grand unified 
groups. Hierarchies and mixings between the second and third generation
are obtained by introducing a single parameter $\epsilon'$
representing the breaking $U(2) \rightarrow U(1)$.
As a consequence, we obtain the successful relations
$ V_{cb} \approx {m_s \over m_b} \approx \sqrt {m_c \over m_t}$.

\vskip 1.0cm

{\bf Acknowledgments}

I am grateful to Lawrence Hall for many valuable comments and
discussions and for constant encouragement. It is also a pleasure to thank 
Zurab Berezhiani, Gia Dvali and Goran Senjanovi\'{c} for some important
remarks, and to K.S. Babu, Biswajoy Brahmachari, Alejandra Melfo, Rabi
Mohapatra and Atsushi Yamada for many useful comments.

\vspace{0.3in}

{\large \bf Appendix}

\vspace{0.2in}

In this Appendix we derive the quark mixing matrices
for the case when vector-like matter with the content of 
$10 + \overline{10}$ of SU(5)
are added to the three generations of the SM
\be
\begin{array}{ccccc}
q_a & u^c_a & d^c_a & l_a & e^c_a \\
Q_1 & U^c_1 & & & E^c_1  \\
Q^c_2 & U_2 & & & E_2 \, .
\end{array}
\ee

The quark mixing matrix is derived as follows
(see also Ref. \cite{lang88}). 
The charged weak current interaction is
\be
L_W \sim [
\overline{{\bf u}}^T \gamma^\mu {\bf d} + 
\overline{U}_1 \gamma^\mu D_1 + 
U^c_2 \gamma^\mu \overline{D^c_2} ] W_\mu^+
\ee

Mass eigenstates are related to the weak eigenstates by
\ba
&{\bf u}_M = {\bf V}^{u\dagger}_L \, 
\left(
\begin{array}{c}
{\bf u} \\
U_1 \\
U_2
\end{array}
\right) \nonumber\\
&{\bf u}^{cT}_M = ( {\bf u}^c \, U^c_1 \, U^c_2) \, {\bf V}^u_R \, ,
\ea
and 
\ba
&{\bf d}_M = {\bf V}^{d\dagger}_L \, 
\left(
\begin{array}{c}
{\bf d} \\
D_1 
\end{array}
\right) \nonumber\\
&{\bf d}^{cT}_M = ( {\bf d}^c \, D^c_2) \, {\bf V}^d_R \, ,
\ea
where $V^u_{L,R}$ and $V^d_{L,R}$ are unitary $5 \times 5$ and
$4 \times 4$ matrices respectively.

The weak current interaction in the mass eigenstate basis is
\ba
L_W & \sim & [
\overline{u}_{Ma} 
\gamma^\mu (V^{u\dagger}_L)_{ai} (V^d_L)_{ib} d_{Mb} \nonumber\\
& + &
u^c_{Ma} \gamma^\mu (V^{u\dagger}_R)_{a5} (V^d_R)_{4b} 
\overline{d^c}_{Mb} \,
] \, W_\mu^+ \, ,
\label{ckm1}
\ea
where $i=1,2,3,4$, $a=1,2,3,4,5$ and $b=1,2,3,4$.

From (\ref{ckm1}) we can read off the quark mixing matrix which is now a 
$10 \times 8$ matrix and consists of two $5 \times 4$ blocks
\ba
V_{a,b} &=& (V^{u\dagger}_L)_{ai} (V^d_L)_{ib}
\nonumber\\
V_{5+a,5+b} &=& (V^{u\dagger}_R)_{a5} (V^d_R)_{4b} 
\ea
and is thus in general {\it not} unitary.
SM mixing matrix is in the upper $3 \times 3$ block of $V_{a,b}$.
The approximate form of V for our model is discussed in the text.

Let us now turn to the neutral current:
\ba
L_Z & \sim & [
\overline{{\bf u}}^T \gamma^\mu {\bf u} + 
\overline{U}_1 \gamma^\mu U_1 -
U^c_2 \gamma^\mu \overline{U^c_2} \nonumber\\
& - & \overline{{\bf d}}^T \gamma^\mu {\bf d} - 
\overline{D}_1 \gamma^\mu D_1 + 
D^c_2 \gamma^\mu \overline{D^c_2}
-j^{\mu}_{em}] Z_\mu
\ea
Now let us go to the mass eigenstate basis.
The electromagnetic part is flavor diagonal. However, the weak part
has flavor changing pieces in the following terms:
\ba
L^{FCNC}_Z & \sim &
\overline{u}^T_{Ma} (V^{u\dagger}_L)_{ai} (V^u_L)_{ic} \gamma^\mu u_{Mc} -
u^c_{Ma} (V^{u\dagger}_R)_{a5} (V^u_R)_{5c} \gamma^\mu \overline{u^c}_{Mc} 
\nonumber\\
& + &
d^c_{Mb} (V^{d\dagger}_R)_{b4} (V^u_R)_{4d} \gamma^\mu \overline{d^c}_{Md} 
\ea
where $i=1,2,3,4$, $a,c=1,2,3,4,5$ and 
$b,d=1,2,3,4$. There 
is no flavor changing part involving left handed down quarks reflecting 
the fact that there is no ``mismatch" in that sector.

From equations (\ref{leftupmix}), (\ref{leftdownmix})
and (\ref{diagup2}), (\ref{diagdown2}) 
we see that in our model the flavor changing Z interactions 
for example for the left-handed up quarks have the following order of
magnitude
\be
( V^{Z-FCNC}_{uu} )_{ac} \equiv
(V^{u\dagger}_L)_{ai} (V^u_L)_{ic} =
\left(
\begin{array}{ccccc}
1 & 0 & 0 & 0 & 0\\
0 & 1 & 0 & {\epsilon' \over \epsilon} {v \over M} & {\epsilon' \over
\epsilon} {v \over M} \\ 
0 & 0 & 1 & {v \over M} & {v \over M} \\ 
0 & {\epsilon' \over \epsilon} {v \over M} & {v \over M} & 1 & 1 \\ 
0 & {\epsilon' \over \epsilon} {v \over M} & {v \over M} & 1 & 1 \\ 
\end{array}
\right)
\ee
These effects are however negligible for M near the GUT scale.

To summarize, there are several consequences of the 
``mismatch":

$\bullet$ The quark mixing matrix is no longer unitary (neither as a 
complete $10 \times 
8$ matrix, neither in the two $5 \times 4$ blocks separately), unless the 
extra $\overline{10}$ is totally decoupled. In particular, it is not
unitary in the $3 \times 3$ standard sector.

$\bullet$ $W_L$ couples also to the ``right-handed" mass eigenstates 
${\bf u}^c_M$ in the lower $5\times 5$ block.

$\bullet$ Couplings of fermions to Z boson are flavor changing.

\end{document}